\documentclass[aps,superscriptaddress,showpacs,floatfix,amsmath,amssymb,nofootinbib,preprintnumbers,
twocolumn]{revtex4-1}

\usepackage{graphicx}

\usepackage{epstopdf}

\usepackage{color}

\newcommand{\be}{\begin{equation}}
\newcommand{\ee}{\end{equation}}
\newcommand{\ba}{\begin{eqnarray}}
\newcommand{\ea}{\end{eqnarray}}

\newcommand{\beq}{\begin{equation}}
\newcommand{\eeq}{\end{equation}}
\newcommand{\beqa}{\begin{eqnarray}}
\newcommand{\eeqa}{\end{eqnarray}}

\newcommand{\bea}{\begin{eqnarray}}
\newcommand{\eea}{\end{eqnarray}}

\begin{document}
\title{Criticality and Surface Tension in Rotating Horizon Thermodynamics}

\author{Devin Hansen}
\email{dhansen@perimeterinstitute.ca}
\affiliation{Perimeter Institute, 31 Caroline St. N., Waterloo,
Ontario, N2L 2Y5, Canada}
\affiliation{Department of Physics and Astronomy, University of Waterloo,
Waterloo, Ontario, Canada, N2L 3G1}
\author{David Kubiz\v n\'ak}
\email{dkubiznak@perimeterinstitute.ca}
\affiliation{Perimeter Institute, 31 Caroline St. N., Waterloo,
Ontario, N2L 2Y5, Canada}
\affiliation{Department of Physics and Astronomy, University of Waterloo,
Waterloo, Ontario, Canada, N2L 3G1}
\author{Robert B. Mann}
\email{rbmann@uwaterloo.ca}
\affiliation{Department of Physics and Astronomy, University of Waterloo,
Waterloo, Ontario, Canada, N2L 3G1}

\date{April 21, 2016}  

\begin{abstract}
We study a modified horizon thermodynamics and the associated criticality for  rotating black hole spacetimes.
Namely, we show that under a virtual displacement of the black hole horizon accompanied by an independent variation of the rotation parameter,
the radial Einstein equation takes a form of a `cohomogeneity two' horizon first law, $\delta E=T\delta S+\Omega \delta J-\sigma\delta A$\,,
where $E$ and $J$ are the horizon energy (an analogue of the Misner--Sharp mass) and the horizon angular momentum, $\Omega$ is the horizon angular velocity, $A$ is the horizon area, and $\sigma$ is the surface tension induced by the matter fields.
For fixed angular momentum, the above equation simplifies and the more familiar (cohomogeneity one) horizon first law
$\delta E=T\delta S-P\delta V$ is obtained, where $P$ is the pressure of matter fields and $V$ is the horizon volume.
A universal equation of state is obtained in each case and the corresponding critical behavior is studied.
\end{abstract}


\maketitle

\section{Introduction}

The thermodynamic interpretation of gravitational field equations has been a subject of intensive study for several decades.
Amongst the various approaches, {\em horizon thermodynamics} \cite{Padmanabhan:2002sha}  provides a very simple and concrete manifestation of the idea that the Einstein equations can be rewritten as a thermodynamic identity  \cite{Cai:2014bwa}.
The original observation was that the radial Einstein equation can be rewritten as a first law of horizon thermodynamics for spherically symmetric black hole spacetimes.  This was subsequently extended  to more general situations
\cite{Padmanabhan:2009vy, Padmanabhan:2013xyr, Kothawala:2007em,Akbar:2007zz, Chakraborty:2015hna}.

For spherically symmetric black hole spacetimes,   horizon thermodynamics essentially identifies
the thermodynamic pressure $P$ with the $T^r{}_r$ component of the energy--momentum tensor of matter fields. 
It is then easy to show that the radial Einstein equation, evaluated on the black hole horizon located at $r=r_+$, can be rewritten as an {\em Horizon Equation of State} (HES)
\be\label{state1}
P=P(V,T)\,,
\ee
which under an infinitesimal virtual displacement of the horizon becomes a (cohomogeneity one)  
{\em Horizon First Law} (HFL)
\be\label{first1}
\delta E=T\delta S-P\delta V\,,
\ee
where $T$ and $S$ are the horizon temperature and entropy identified through standard thermodynamic arguments. The quantity  $E=r_+/2$ is an horizon energy (equal to the Misner--Sharp energy evaluated on the horizon), and $V$ is the geometric volume associated with the black hole,
\be
P=T^r{}_r|_{r+}\,,\quad V=\frac{\Sigma_{d-2} r_+^{d-1}}{d-1}\,,
\ee
where for $d$ spacetime dimensions, $\Sigma_{d-2}$ denotes a finite volume of the $(d-2$)-dimensional `unit sphere'.
We emphasize that both horizon equations \eqref{state1} and \eqref{first1} are {\em universal}, that is, independent of matter content and
entirely fixed by the gravitational theory under consideration. (The actual matter dependence enters entirely through the pressure term $P$.)

Interestingly, similar equations have been recently studied for asymptotically AdS black holes in the
context of {\em extended phase space thermodynamics}, e.g. \cite{KastorEtal:2009, Dolan:2014jva, Kubiznak:2014zwa, Altamirano:2014tva}, where
one identifies the thermodynamic pressure $P_\Lambda$ with the (negative) cosmological constant $\Lambda$ (which is allowed to vary in the first law)
and defines  thermodynamic volume $V_{\mbox{\tiny TD}}$ as the quantity thermodynamically conjugate to $P_\Lambda$.
For stationary 
black hole spacetimes in Einstein gravity with angular momentum ${\cal J}$ and horizon angular velocity $\Omega$ that are
coupled to $U(1)$ charge $Q$ (with a corresponding chemical potential $\Phi$)  this results in the following equation of state and the extended first law:
\ba\label{Ext1st}
P_\Lambda&=&P_\Lambda(V_{\mbox{\tiny TD}}, T, Q, {\cal J})\,,\nonumber\\
\delta M&=& T\delta S+ V_{\mbox{\tiny TD}}\delta P_\Lambda +\Phi \delta Q+\Omega\delta {\cal J}\,,
\ea
where $M$ stands for the black hole mass interpreted now as a gravitational enthalpy \cite{KastorEtal:2009},
and
\be
P_\Lambda=-\frac{\Lambda}{8\pi}\,,\quad V_{\mbox{\tiny TD}}=\Bigl(\frac{\partial M}{\partial P_\Lambda}\Bigr)_{S,Q, {\cal J}}\,.
\ee
Contrary to \eqref{first1}, the extended first law \eqref{Ext1st} is typically of maximal cohomogeneity.
The two approaches were, {\em for spherically symmetric spacetimes}, 
compared in \cite{Ma:2015llh, Hansen:2016ayo}, where  universality of the $P-V$ criticality of the horizon equation of state \eqref{state1} was also demonstrated.

Surprisingly, despite its relative success for black holes with spherical symmetry, horizon thermodynamics has not really been fully extended to rotating black hole spacetimes. Indeed, only a few studies exist in this direction. It was shown in \cite{Kothawala:2007em} for the Kerr-Newmann case and in \cite{Akbar:2007zz} for the charged BTZ black hole that the radial Einstein equation can be rewritten as a `standard' thermodynamic first law.  We  critically comment on 
the corresponding procedure, which explicitly uses the properties of a given solution, in  App.~B.  An implicit study of the rotating case is also contained in \cite{Chakraborty:2015hna} where an HFL \eqref{first1} is obtained for an arbitrary null surface.  However, in all  studies of horizon thermodynamics so far 
the obtained first law is only of cohomogeneity one. The only admissible variation is due to the virtual displacement of the horizon, which clearly neglects additional features that are present due to rotation.

The aim of this paper is to formulate an extension of horizon thermodynamics to  rotating black hole spacetimes that has the following `natural' features: i) it maintains the `universality' regarding the matter content of the theory, that is, no particular  properties of a given solution to the field equations are exploited and ii) additional degrees of freedom associated with   rotation are properly captured and allowed to vary in the HFL, leading naturally to a description of higher rank cohomogeneity.

As we shall see for a given ansatz for the rotating black hole geometry below, see Eq.~\eqref{metric}, the radial Einstein equation naturally results in the following modified HES and HFL:\footnote{ In fact, the HES \eqref{state2} and the HFL \eqref{first2} are more general than their derivation employing the ansatz \eqref{metric} may suggest---we expect them to equally hold for asymptotically flat rotating black holes in higher dimensions, or, as demonstrated in App.~\ref{AppAdS}, for black holes with a cosmological constant.}
\ba
\sigma&=&\sigma(A,T, J)\,,\label{state2}\\
\delta E&=&T\delta S+\Omega \delta J-\sigma \delta A\,,\label{first2}
\ea
where $E$ is the horizon energy analogous to the Misner--Sharp mass, $J$ and $\Omega$ are the horizon angular momentum and velocity,
$A$ is the horizon area, and $\sigma$ is the {\em surface tension} induced by the matter fields.

Contrary to the first law \eqref{first1}, the modified HFL is a cohomogeneity two identity, as both the
horizon radius $r_+$ and the associated rotation parameter $a$ are allowed to vary independently.
However, although universal in the sense  that all matter dependence is encapsulated in the surface
tension $\sigma$, the modified HFL \eqref{first2} depends crucially on the applicability of the given metric ansatz. (Only a limited number of solutions to Einstein-matter equations can be written in the form we consider.) This is in strong contrast to the spherically symmetric case, in which metric ansatz is general (within the symmetry requirements)
and the horizon first law is completely universal,  dependent only on the type of gravitational theory but not its matter content \cite{Hansen:2016ayo}.

Finally, we show that in the case when an additional horizon structure---a black hole volume $V$---is independently identified, the HES \eqref{state2} and the HFL \eqref{first2} can be rewritten in a more familiar form
\ba
P&=&P(V,T,J)\,,\label{state3}\\
\delta E&=&T\delta S+\Omega \delta J-P\delta V\,,\label{first3}
\ea
where pressure $P$ depends on the matter content and is defined as a quantity thermodynamically conjugate to $V$. However, to obtain these relations, either a very specific form of the volume has to be considered, or one has to restrict to a fixed  angular momentum ensemble 
and accept the consequence that the HFL \eqref{first3} is only a cohomogeneity one equation.

Our paper is organized as follows. In the next section we derive the modified HES and HFL \eqref{state2} and \eqref{first2}, and
study their criticality. 
In Sec.~4 an additional structure of horizon volume is assumed and the horizon equations are rewritten in the more familiar (generically cohomogeneity one) form  \eqref{state3} and \eqref{first3}. 
Sec.~5 is devoted to discussion and conclusions. App.~A describes an alternate derivation of horizon equations \eqref{state2} and \eqref{first2} and
App.~\ref{AppAdS} demonstrates that these equations also apply to black holes with cosmological constant. Finally,
in App.~C we critically revise an argument showing the `equivalence' of the radial Einstein equation with the standard first law of black hole thermodynamics for the Kerr--Newmann black hole.

\section{Rotating horizon thermodynamics}

For concreteness and simplicity we work with the following ansatz for a rotating black hole geometry
\ba\label{metric}
ds^2&=&-\frac{\Delta}{\rho^2}\bigl(dt-a\sin^2\!\theta d\varphi\bigr)^2+\frac{\rho^2}{\Delta}dr^2
\nonumber\\
&&+\rho^2d\theta^2+\frac{\sin^2\!\theta}{\rho^2}\bigl[adt-(r^2+a^2)d\varphi\bigr]^2\,,
\ea
generalizing the Kerr metric, where
\be
\rho^2=r^2+a^2\cos^2\!\theta\,,
\ee
and we assume that the metric function $\Delta=\Delta(r)$ determines the position of the (non-extremal) black hole horizon located at the largest root of $\Delta(r_+)=0$.

\subsection{Modified horizon equations}

We begin by deriving the modified HFL and HES \eqref{state2} and \eqref{first2}, assuming   Einstein gravity minimally coupled to matter. 
From the geometry we can immediately identify the black hole horizon area
\be\label{A}
A=4\pi (r_+^2+a^2) = 4S\,,
\ee
in terms of the entropy $S$. The horizon angular velocity
\be\label{Omega}
\Omega=-\frac{g_{t\varphi}}{g_{\varphi\varphi}}\bigg|_{r_+}=\frac{a}{r_+^2+a^2}
\ee
can likewise be identified, as can the black hole temperature
\be\label{temp1}
T=\frac{\Delta'(r_+)}{4\pi(r_+^2+a^2)}\,,
\ee
via standard Wick-rotation arguments. Note that no field equations are required up to this point, though
the latter relation in \eqref{A}  employs the assumption of Einstein gravity.

Let us next consider the radial Einstein equation, evaluated on the black hole horizon
\be
8\pi T^r{}_r|_{r_+}=G^r{}_r|_{r_+}=\frac{a^2-r_+^2+r_+\Delta'(r_+)}{\rho_+^4}\,,
\ee
where $\rho_+^2=r_+^2+a^2\cos^2\!\theta$.
Using \eqref{temp1} we obtain
\be\label{temp2}
T=\frac{8\pi \rho_+^4 T^r{}_r|_{r_+} +r_+^2-a^2}{4\pi r_+(r_+^2+a^2)}\,,
\ee
which yields
\be\label{ho}
T\delta S=\frac{2\rho_+^4 T^r{}_r|_{r_+}}{r_+(r_+^2+a^2)}\delta S+\frac{r_+^2-a^2}{4\pi r_+(r_+^2+a^2)}\delta S
\ee
upon multiplication  by $\delta S=2\pi (r_+\delta r_++a \delta a)$. Note that the  first term on the right-hand-side of
\eqref{ho}  depends on the matter content, whereas the second term is universal and completely
fixed in terms of $r_+$ and $a$.

Now we make the following interesting observation. This latter  term in \eqref{ho} can be written as
\be\label{ho2}
\frac{r_+^2-a^2}{4\pi r_+(r_+^2+a^2)}\delta S=\delta E-\Omega \delta J\,,
\ee
upon defining
\be\label{EJ}
E=\frac{r_+^2+a^2}{2r_+}\,, \quad J=Ea\,,
\ee
the former quantity being defined only up to a total variation.
We see that the expressions for $E$ and $J$ are formally identical to those for the mass and angular momentum of vacuum Kerr black hole, respectively. Furthermore,  in the absence of rotation, $a\to0$,  $E$ reduces to the Misner--Sharp energy of a spherically symmetric spacetime evaluated on the black hole horizon.  We therefore identify $E$ as the {\em horizon energy} $E$ and $J$ as the {\em horizon angular momentum} of the black hole described by \eqref{metric}.

We have thus found the following relation:
\be
\delta E=T\delta S+\Omega \delta J-\frac{2\rho_+^4T^r{}_r|_{r_+}}{r_+(r_+^2+a^2)}\delta S\,.
\ee
Since $T$ must be constant on the horizon \cite{Medved:2004tp}, the last
equation is consistent only when $\rho_+^4 T^r{}_r|_{r_+}$ is independent of coordinate $\theta$.
We therefore introduce the surface tension
\be\label{sigma}
\sigma=\sigma(r_+,a)=\frac{\rho_+^4T^r{}_r|_{r_+}}{2r_+(r_+^2+a^2)}\,,
\ee
and so obtain \eqref{first2} for the modified HFL
\be\label{first4}
\delta E=T\delta S+\Omega \delta J-\sigma \delta A\,.
\ee

We pause to comment that a surface tension $\sigma$ and a first law of the type $\delta E=T\delta S-\sigma\delta A$ were first considered by York \cite{York:1986it} in the context of the thermodynamics of black holes in a cavity. However there is a fundamental difference. York's `surface tension' is conjugate to the area of a cavity enclosing the ensemble and so the cavity area $A$ and the black hole entropy $S$ can vary independently.  In our case we have no cavity. Instead the surface tension $\sigma$ is conjugate to the area of black hole horizon itself and is entirely induced by the matter fields present in the spacetime. In particular, in vacuum we have $\sigma=0$ and recover the standard 1st law of black hole thermodynamics
\be
\delta E=T\delta S+\Omega \delta J\,,
\ee
whereas in the electrovacuum (Kerr--Newman) case, we have
\be
T^r{}_r|_{r_+}=-\frac{Q^2}{8\pi \rho_+^4}\,\quad \Rightarrow \quad
\sigma=-\frac{Q^2}{16\pi r_+(r_+^2+a^2)}\,.
\ee

The HFL \eqref{first4} is cohomogeneity two as both the horizon radius $r_+$ and the rotation parameter $a$ can vary independently.
Moreover, Eq.~\eqref{temp2} together with \eqref{sigma} yields
\be\label{state4}
\sigma=\sigma(A,J,T)=
\frac{T}{4}+\frac{a^2-r_+^2}{16\pi r_+(r_+^2+a^2)}\,,
\ee
which is the  surface tension HES \eqref{state2}.
Here $r_+$ and $a$ are implicitly given in terms of $J$ and $A$ through relations \eqref{EJ} and \eqref{A}.

Equations \eqref{first4} and \eqref{state4} are together with the definition of the surface tension \eqref{sigma} the most important results of this section. Note that in order to write these equations down, no new quantities, apart from $E$ and $J$, had to be defined and the expressions are entirely given in terms of geometric horizon properties such as the area $A$, temperature $T$, and angular velocity $\Omega$.

.

\subsection{Surface tension criticality} 

\begin{figure}
\begin{center}
\includegraphics[width=0.47\textwidth,height=0.31\textheight]{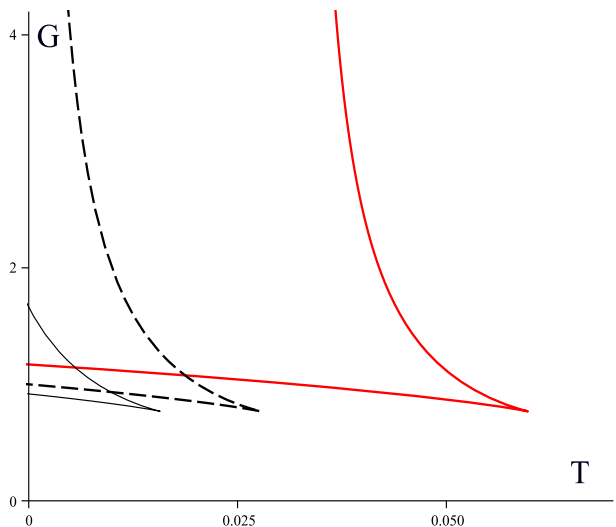}
\caption{ {\bf $\sigma-A$ criticality.} The $G_\sigma-T$ diagram is displayed for $J=1$ and various $\sigma$. The red curve corresponds to positive tension $\sigma=0.008$, the dashed black curve to $\sigma=0$, and the thin black curve to negative tension $\sigma=-0.003$. We observe a characteristic cusp whose position depends on $\sigma$. For positive $\sigma$, the correspponding upper branch terminates at finite temperature $T$.
}
\label{fig1}
\end{center}
\end{figure}
Let us now study the possible critical behavior associated with the generalized horizon thermodynamics derived in the previous subsection. For concreteness, we do this in a canonical (fixed $J$) ensemble.

Since according to the HFL \eqref{first4}, the quantity $E$ in \eqref{EJ} plays the role of thermodynamic energy (that is a thermodynamic potential expressed in terms of extensive thermodynamic variables $S, J$ and $A$), we define
\be
G_\sigma=G_\sigma(T,\sigma, J)=E-TS+\sigma A\,,
\ee
which is the corresponding surface tension Gibbs free energy $G_\sigma$.
This quantity formally satisfies 
\be
\delta G= -S\delta T+\Omega \delta J+A\delta \sigma\,.
\ee
The behavior of $G=G(T, \sigma, J)$ is displayed in Fig.~\ref{fig1} for fixed $J=1$ and three representative values of $\sigma$.
For any $\sigma$ we observe two branches of black holes, meeting at a characteristic cusp. For negative $\sigma$ both branches on the other end terminate at finite $G$ and $T=0$, whereas for positive $\sigma$ the upper branch eventually asymptotes to $G\to \infty$ at  $T=4\sigma$, with a divergence at $T=0$ occurring for $\sigma=0$.  Apart from the presence of a cusp, no interesting thermodynamic behaviour is observed  for any values of $J$.

As with the spherically symmetric case \cite{Hansen:2016ayo}, an interpretation of  the concrete thermodynamic behaviour depends on the actual matter content. For example, in vacuum, $\sigma=0$ and only  the black dashed curve applies.
Similarly, for the electrovacuum case with nontrivial charge $\sigma<0$ and  behavior similar to the thin black curve in Fig.~\ref{fig1} is realized.  We expect that our ansatz could be suitably generalized to accomodate
rotating black hole with some type of a scalar hair \cite{Herdeiro:2015waa}, with free energy plots similar to
the positive $\sigma$ curve.

\subsection{Effective temperature}
\begin{figure}
\begin{center}
\includegraphics[width=0.47\textwidth,height=0.31\textheight]{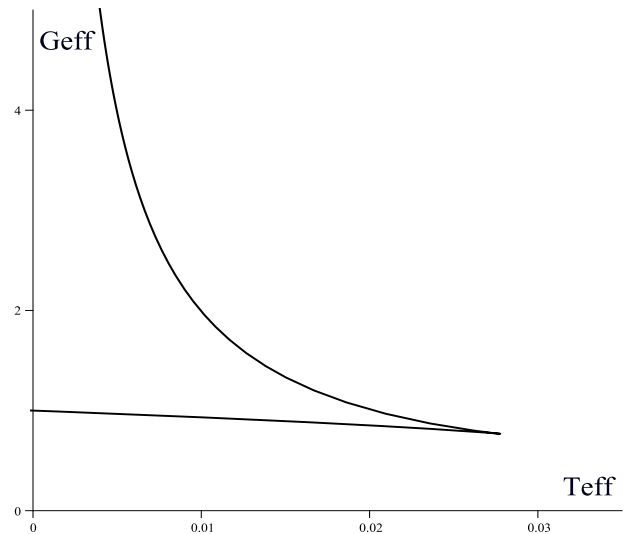}
\caption{ {\bf Universal criticality.} The $G_{T_{\mbox{\tiny eff}}}-{T_{\mbox{\tiny eff}}}$ phase diagram is displayed for $J=1$.
We observe a characteristic cusp that is completely independent of the matter content of the theory.
}
\label{fig2}
\end{center}
\end{figure}

The modified HFL  \eqref{first4} has three terms on its right-hand-side but inherently   is only cohomogeneity two. Furthermore, variation of $S$ is not independent of the variation of $A$. This suggests that we
introduce an {\em effective temperature}
 \be
T_{\mbox{\tiny eff}}=T-4\sigma =\frac{1}{4\pi}\frac{r_+^2-a^2}{r_+(r_+^2+a^2)}\,,
\ee
which is easily obtained by  grouping the $T\delta S$ and $-\sigma \delta A$ terms together. Note that this quantity has no explicit dependence on matter, is constant on the horizon, and is positive for $r_+>a$.
With this identification, the modified HFL \eqref{first4} becomes manifestly of cohomogeneity two and reads
\be\label{effective}
\delta E=T_{\mbox{\tiny eff}}\delta S+\Omega \delta J\,,
\ee
which is equivalent to Eq.~\eqref{ho2}. In fact, since $E$ and $J$ coincide with the mass and angular momentum of the vacuum Kerr black hole,
the effective temperature $T_{\mbox{\tiny eff}}$ is nothing  other  than the temperature of the Kerr solution and  \eqref{effective} is the corresponding first law.

Stated  this way, horizon thermodynamics is recast in universal form that is completely independent of the matter content and
represented by the thermodynamics of a vacuum solution. Note that the same is true in the case of spherical symmetry upon absorbing  the
$-PdV$ term into $TdS$ in \eqref{first1}, which then simply reads $\delta E=T_{\mbox{\tiny eff}}\delta S$, with $T_{\mbox{\tiny eff}}$ being the temperature of the Schwarzschild black hole.
This interpretation of horizon thermodynamics also opens a new way of deriving the horizon equations \eqref{first4} and \eqref{state4},  as we demonstrate in App.~A.

 In the light of previous discussion, it is obvious that the criticality of the HFL \eqref{effective} coincides with that of the Kerr solution. Namely, the associated Gibbs free energy reads
\be
G_{T_{\mbox{\tiny eff}}}=G_{T_{\mbox{\tiny eff}}}({T_{\mbox{\tiny eff}}}, J)=E-{T_{\mbox{\tiny eff}}}S
=\frac{r_+^2+3a^2}{4r_+}\,,
\ee
and obeys
\be
\delta G_{T_{\mbox{\tiny eff}}}=-S\delta {T_{\mbox{\tiny eff}}}+\Omega \delta J\,.
\ee
The corresponding $G_{T_{\mbox{\tiny eff}}}=G_{T_{\mbox{\tiny eff}}}({T_{\mbox{\tiny eff}}}, J)$ diagram is displayed in Fig.~\ref{fig2}.
For non-trivial angular momentum $J$, we observe a characteristic cusp,  completely independent of the matter content of the theory.

To summarize this section, we  stress that both the surface tension and the effective temperature approaches are very natural in the horizon thermodynamics of rotating black holes. Both permit study of cohomogeneity two HFLs since variations of both $\delta a$ and $\delta r_+$ are allowed. Furthermore,  there is no need to identify any extra structure beyond the horizon energy $E$ and angular momentum $J$ in \eqref{EJ}.
We shall now consider an alternate approach in which
an additional structure, the black hole volume $V$, is defined.

\section{$P-V$ criticality}
\label{volsect}

We now consider the implications of rewriting the last term in \eqref{first4} as a pressure-volume term
\be\label{what}
\sigma \delta A=P\delta V\,,
\ee
an equality that is only possible when $\delta V\propto \delta A$.
Specifically we shall consider two approaches.  Motivated by scaling properties, we consider $V\propto A^{3/2}$, which
yields cohomogeneity two HFL in which $a$ and $r_+$ can be independently varied.  We then consider the alternate possibility in which
\be
a=a(r_+)\,,
\ee
while $V$ is `independently specified' by other criteria, e.g. identified with the geometric/thermodynamic volume of the black hole.

\subsection{Volume as power of area}

To keep the HFL of cohomogeneity two we impose\footnote{Note that such a
 `power law rewriting' is similar in spirit to what has recently been discussed in \cite{Armas:2015qsv} in the context of extended phase space thermodynamics.
}
\be\label{V1}
V=\frac{4}{3}\pi\left(\frac{A}{4\pi}\right)^{3/2}\,,
\ee
where the constant of proportionality has been chosen to yield $V=\frac{4}{3}\pi r_+^3$ in the limit of zero rotation \cite{Hansen:2016ayo}.
This then
yields
\ba
P&=&P(V,T,J)\,,\label{state5}\\
\delta E&=&T\delta S+\Omega \delta J-P\delta V\label{first5}
\ea
for the HES and HFL, respectively.
The associated Gibbs free energy is defined through
\be
G_P = E - TS + PV\,.
\ee
It obeys
\be
\delta G_P=-S\delta T+\Omega \delta J+V\delta P\,,
\ee
and encodes information about possible thermodynamic phase transitions.

Specifically, the identification \eqref{what} implies
\be
P=\frac{\sigma A}{2\pi }\Bigl(\frac{4\pi}{A}\Bigr)^{3/2}\,
\ee
for the pressure. Upon using \eqref{state4},   the HES \eqref{state5} then becomes
\be
P=\frac{T}{2\sqrt{r_+^2+a^2}}+\frac{a^2-r_+^2}{8\pi r_+(r_+^2+a^2)^{3/2}}\,,
\ee
where $r_+$ and $a$ are implicit functions of $V$ and $J$, according to \eqref{V1} and \eqref{EJ}.

Note that in the limit $a\to 0$ we recover that $P=T^r{}_r|_{r_+}$, as identified in
the spherically symmetric scenario \cite{Hansen:2016ayo}, and the HES reduces to the HES for spherical black holes in
Einstein gravity
\be\label{EosEin}
P=\frac{T}{2r_+}-\frac{1}{8\pi r_+^2}\,.
\ee
(The same will be true for the HESs derived in the next subsection.)

\begin{figure}
\begin{center}
\includegraphics[width=0.47\textwidth,height=0.31\textheight]{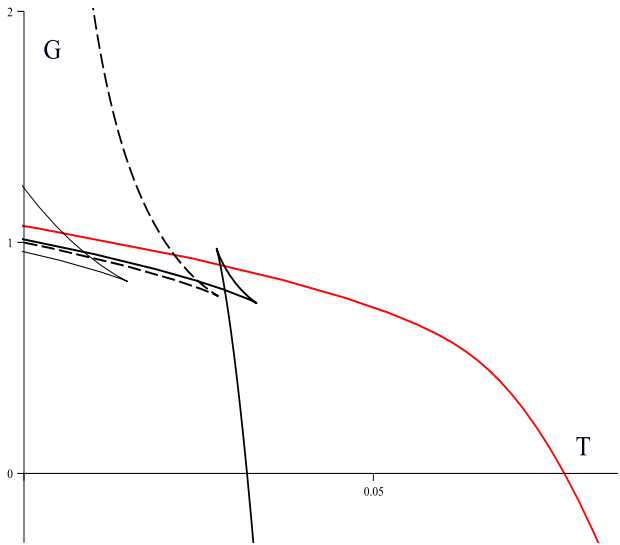}
\caption{\textbf{$P-V$ criticality: $V=\frac{4}{3}\pi[A/(4\pi)]^{3/2}$ case.} The $G_P-T$ diagram is displayed for $J=1$ and fixed pressure. Namely, the red curve corresponds to positive pressure $P=0.007$, the thick black curve to $P=0.003$ and shows the characteristic swallow tail, the dashed black curve to $P=0$, and the thin black curve to negative $P=-0.003$.}
\label{Fig6}
\end{center}
\end{figure}
The corresponding $G-T$ diagram is displayed in Fig.~\ref{Fig6}. We observe that for positive $P<P_c$ there is a characteristic swallow tail indicating the presence of Van der Waals-like phase transition, e.g. \cite{Altamirano:2014tva}.

\subsection{Fixed angular momentum}

Let us now study the situation when variations $\delta a$ and $\delta r_+$ are not independent.
Specifically, let us consider the
canonical ensemble where the horizon angular momentum is held fixed, $\delta J=0$.
This implies that $a=a(r_+)$ and\footnote{Note that it is a standard practice in horizon thermodynamics that only one thermodynamic parameter is allowed to vary while other parameters are held constant. Namely, one is allowed to ``virtually displace the horizon'', changing $r_+$, \cite{Padmanabhan:2002sha,Padmanabhan:2009vy, Padmanabhan:2013xyr, Kothawala:2007em, Akbar:2007zz, Chakraborty:2015hna}.
In this subsection we essentially return back to a single parameter variation. However, contrary to the `standard' practice, 
we consider `induced' variation, $\delta a=\delta a(r_+)$, as required by the canonical ensemble.
}
\be
\delta a=\frac{a(a^2-r_+^2)}{r_+(r_+^2+3a^2)}\delta r_+\,.
\ee
Upon using \eqref{what}, the HES and the HFL take the form
\ba
P&=&P(V,T)\,,\\
\delta E&=&T\delta S-P\delta V\,.
\ea
Novel feature in these expressions is that the black hole volume is a new thermodynamic quantity that is independently specified by some other criteria. In what follows we consider two examples: i) the geometric volume of the Kerr black hole and ii) the thermodynamic volume of the Kerr black hole, e.g. \cite{Cvetic:2010jb}.

\begin{figure}
\begin{center}
\includegraphics[width=0.47\textwidth,height=0.31\textheight]{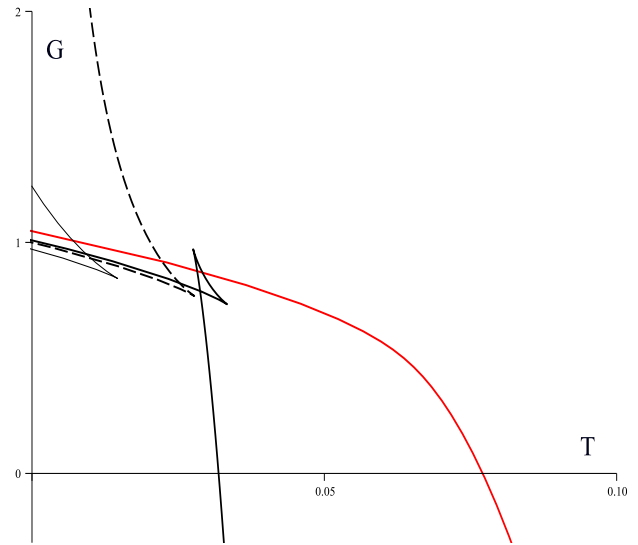}
\caption{\textbf{$P-V$ criticality: $V=r_+A/3$ case.} The $G-T$ diagram is displayed for $J=1$ and fixed pressure. The behavior is qualitatively similar to that in Fig.~\ref{Fig6}. Namely, the red curve corresponds to small positive pressure $P=0.007$, the thick black to $P=0.0012$ and shows the characteristic swallow tail, the dashed black to $P=0$ and the thin black to negative $P=-0.003$. (Note that the pressures are 10 times magnified relative to the previous figure.)}
\label{Fig7}
\end{center}
\end{figure}

Considering first the geometric volume
\be\label{V2}
V=\frac{r_+A}{3}\,,
\ee
Eq.~\eqref{what} yields
\be
P=\frac{6\sigma(r_+^2+a^2)}{r_+(3r_+^2+5a^2)}\,,
\ee
which upon using \eqref{state4} becomes
\be
P=\frac{3}{2}\frac{T(r_+^2+a^2)}{r_+(3r_+^2+5a^2)}+\frac{3}{8\pi}\frac{a^2-r_+^2}{r_+^2(3r_+^2+5a^2)}\,,
\ee
where $a$ and $r_+$ are implicit functions of $V$ and $J$ through \eqref{V2} and \eqref{EJ}.
The corresponding behavior of $G=E-TS+PV=G(P,T)$, for various values of fixed $P$ is illustrated in Fig.~\ref{Fig7} and is qualitatively similar to
behavior displayed in Fig.~\ref{Fig6}.

Another choice, motivated by extended phase space thermodynamics \cite{Altamirano:2014tva} is to
set $V$ equal to  the thermodynamic volume of Kerr black hole,
\be
V=\frac{r_+A}{3}\Bigl(1+\frac{a^2}{2r_+^2}\Bigr)\,,
\ee
in which case we recover,
\be
P=\frac{12\sigma r_+(r_+^2+a^2)}{6r_+^4+9r_+^2a^2+a^4}\,,
\ee
or
\be
P=\frac{3Tr_+(r_+^2+a^2)}{6r_+^4+9r_+^2a^2+a^4}+\frac{3}{4\pi}\frac{a^2-r_+^2}{6r_+^4+9r_+^2a^2+a^4}\,,
\ee
upon using \eqref{state4}.
The corresponding $G=G(P,T)$ behavior is qualitatively similar to that displayed in Fig.~\ref{Fig7}. 


\section{Summary}

 We have extended horizon thermodynamics from its traditional spherically symmetric ansatz~\cite{Padmanabhan:2002sha} 
 to  rotating black hole spacetimes.
 The horizon area, temperature, and horizon angular velocity of the black hole are all straightforwardly identified.  However to rewrite the Einstein equations for the axially symmetric metric ansatz \eqref{metric} as a first law of horizon thermodynamics it is necessary to make identifications for the angular momentum $J$ and energy $E$.  These can be well motivated from the radial Einstein equation,  but they are not uniquely defined (see however the discussion in App.~A). We expect that provided these quantities are appropriately re-defined, the same HFL \eqref{first2} will hold for other rotating black holes, not necessarily described by the ansatz \eqref{metric}, including black holes in higher dimensions or those with cosmological constant (see App.~\ref{AppAdS}).

Our approach goes beyond traditional horizon thermodynamics, where one varies only the horizon radius. Our construction allows for both the horizon radius and the angular momentum to vary independently of one another, thus allowing for a fuller spectrum of thermodynamic possibilities.
This involved identifying a ``surface tension" $\sigma$, writing the work term as $\delta W = -\sigma \delta A$.  This is a provocative identification, as the term has the correct units and is dependent on matter content.
However, any further reason for making such an identification remains elusive  at the moment and will be a subject of further study.

We have also shown that horizon thermodynamics  naturally identifies an effective temperature  $T_{\mbox{\tiny eff}} = T + T_m$, where $T_m = -4\sigma$.  This relationships follows from recognizing that the  ``work'' term $\delta W = -\sigma \delta A = T_m \delta S$.  In other words Hawking temperature $T$ of a solution is a difference between the vacuum Hawking temperature $T_{\mbox{\tiny eff}}$ and the matter field contribution $T_m$.  Making this identification yields a universal  first law that is independent of matter content.

While it is tempting to see if we can eliminate the notion of surface tension in terms of pressure by introducing the notion of volume, we  have found that this is somewhat problematic.  As we demonstrated in Sec.~\ref{volsect},
 this entails either a very specific choice of the black hole volume or restricting to the canonical ensemble, yielding
a cohomogeneity-one first law.  In the latter case the volume can be freely specified by other criteria and is not restricted by horizon thermodynamics.
This stands in stark contrast to the extended phase space approach, in which volume is uniquely defined as the conjugate of thermodynamic pressure \cite{KastorEtal:2009, Dolan:2014jva, Kubiznak:2014zwa, Altamirano:2014tva}, the latter being proportional to a cosmological constant.

\section*{Acknowledgements}
We are grateful to the anonymous referee for the valuable comments on our manuscript.
This research was supported in part by Perimeter Institute for Theoretical Physics and by the Natural Sciences and Engineering Research Council of Canada. Research at Perimeter Institute is supported by the Government of Canada through Industry Canada and by the Province of Ontario through the Ministry of Research and Innovation.

\appendix

\section{Alternate derivation of horizon equations}

Although precisely in the spirit of horizon thermodynamics \cite{Padmanabhan:2002sha} generalized to the rotating case, the derivation of the horizon equations \eqref{first4} and \eqref{state4} in the main text suffers from non-uniqueness of the definition of horizon energy $E$ and horizon angular momentum $J$, \eqref{EJ}. Although their definition is motivated by \eqref{ho2}, the possibility of redefining $E$ by a total derivative
(accompanied by a proper modification of $J$) remains. For this reason in this appendix we give an alternate derivation of these equations, turning around the logic of the reasoning. Namely, we start again with the ansatz \eqref{metric} but consider the vacuum solution first.\footnote{This goes directly against the spirit of horizon thermodynamics that essentially tries to avoid working with concrete solutions of field equations.}
This allows us to identify $E$ and $J$. We then carry the analysis to the non-vacuum case, keeping the same $E$ and $J$ to rederive Eqs. \eqref{first4} and \eqref{state4} in a different fasion.

Let us start again with the ansatz \eqref{metric} and specify to the vacuum Kerr case, setting $\Delta=r^2-2mr+a^2$. The thermodynamic quantities then read
\ba\label{A1}
E&=&m=\frac{r_+^2+a^2}{2r_+}\,,\quad J=Ea\,,\nonumber\\
\Omega&=&\frac{a}{a^2+r_+^2}\,,\quad S=\frac{A}{4}=\pi(r_+^2+a^2)\,,\nonumber\\
T_0&=&T_{\mbox{\tiny eff}}=\frac{r_+^2-a^2}{4\pi r_+(r_+^2+a^2)}\,,
\ea
and obey the standard first law
\be\label{A2}
\delta E=T_0\delta S+\Omega \delta J\,,
\ee
which is of course identical to the effective first law \eqref{effective}.

We next consider the spacetime with matter, keeping the same ansatz \eqref{metric} and general $\Delta=\Delta(r)$ that determines the position of the horizon. 
The derivation of the horizon equations \eqref{first4} and \eqref{state4} then consists of the following 4 steps:
\begin{itemize}
\item
We insist that even in the presence of mater the horizon energy $E$ and the horizon angular momentum $J$ are given by the vacuum expressions \eqref{A1}. (This in some sense directly generalizes the idea of Misner--Sharp quantities to the case with rotation.) 
\item
We employ the Euclidean trick to identify the actual temperature of the black hole horizon according to
\be
T=\frac{\Delta'(r_+)}{4\pi(r_+^2+a^2)}\,.
\ee
\item
We impose the radial Einstein equation evaluated on the horizon, to relate $T$ and $T_0$,
\be
\qquad 8\pi T^r{}_r|_{r_+}=G^r{}_r|_{r_+}=\frac{a^2-r_+^2+r_+\Delta'(r_+)}{\rho_+^4}
\ee
\bigskip
which rewrites, upon using \eqref{A1}, as
\be
T=T_0+4\sigma\,,\quad \sigma\equiv \frac{8\pi \rho_+^4T^r{}_{r}|_{r=r_+}}{4\pi r_+(r_+^2+a^2)}\,.
\ee
So we identified the matter contribution to the temperature called surface tension $\sigma$ in the main text,
and recovered the HES \eqref{state4}.
\item
The final step is to rewrite the standard first law \eqref{A2} in terms of the actual temperature in the presence of matter,
\be
\delta E=T_0\delta S+\Omega \delta J=T\delta S+\Omega \delta J-\sigma \delta A\,,
\ee
which is the HFL \eqref{first4}.
\end{itemize}

 We believe that this derivation in some sense reveals the true nature of horizon thermodynamics. It describes the
standard vacuum first law from a perspective of an observer who measures the actual black hole temperature $T$ and the
surface tension $\sigma$ associated with matter fields present in the spacetime. This is the origin of universality of horizon
thermodynamics: all black holes satisfy `an equivalence class' of first laws \eqref{A2} irrespective of the matter content of the
theory. Specific features of a given black hole emerge only after the actual matter content and associated conserved charges are identified, along with their respective contributions to the first law.

 Of course, exactly the same derivation would apply to the spherically symmetric case.

\section{Asymptotically AdS rotating horizon thermodynamics}\label{AppAdS}

Let us now show that we recover the same HFL \eqref{first2} also in the asymptotically AdS case, showing hence that
the result is more general than `its derivation' through the ansatz \eqref{metric}.

To this end we consider a more general ansatz for an asymptotically AdS rotating black hole geometry
given by
\ba\label{metric}
ds^2&=&-\frac{\Delta}{\rho^2}\bigl(dt-a\sin^2\!\theta \frac{d\varphi}{\Xi}\bigr)^2+\frac{\rho^2}{\Delta}dr^2
\nonumber\\
&&+\rho^2d\theta^2+\frac{\Sigma\sin^2\!\theta}{\rho^2}\bigl[adt-(r^2+a^2)\frac{d\varphi}{\Xi}\bigr]^2\,,
\ea
generalizing the Kerr-AdS metric, where
\ba
S&=&1-\frac{a^2}{l^2}\cos^2\!\theta\,,\quad \Xi=1-\frac{a^2}{l^2}\,,\nonumber\\
\rho^2&=&r^2+a^2\cos^2\!\theta\,,
\ea
and we assume that the metric function $\Delta=\Delta(r)$ determines the position of the (non-extremal) black hole horizon located at the largest root of $\Delta(r_+)=0$. To ensure the AdS asymptotics we further assume the following large-$r$ expansion:
\be\label{asympt}
\Delta=\frac{r^4}{l^2}+o(r^4)\,.
\ee

We now follow the same procedure as for the asymptotically flat case. We identify the entropy
\be\label{S2}
S=\frac{A}{4}=\pi \frac{r_+^2+a^2}{\Xi}\,,
\ee
and the horizon angular velocity
\be\label{Omega}
\Omega_H=-\frac{g_{t\varphi}}{g_{\varphi\varphi}}\bigg|_{r_+}=\frac{a\Xi}{r_+^2+a^2}\,.
\ee
Since the solution is written in rotating coordinates, one has to subtract the rotation at infinity. Using \eqref{asympt}, we find $\Omega_{\infty}=-a/l^2$, and hence the angular velocity that enters the first law is
\be
\Omega=\Omega_H-\Omega_{\infty}=\frac{a(r_+^2+l^2)}{l^2(r_+^2+a^2)}\,.
\ee
We also identify the black hole temperature
\be\label{temp1}
T=\frac{\Delta'(r_+)}{4\pi(r_+^2+a^2)}\,.
\ee
Employing the radial Einstein equation, evaluated on the black hole horizon, we have
\ba
8\pi T^r{}_r|_{r_+}&=&\bigl(G^r{}_r-\frac{3}{l^2}\delta^r_r\bigr)|_{r_+}\nonumber\\
&=&\frac{1}{r\rho_+^4}\Bigl[a^2-r_+^2+r_+\Delta'(r_+)-\frac{3r_+^4}{l^2}-\frac{r_+^2a^2}{l^2}\Bigr]\,,\nonumber
\ea
where $\rho_+^2=r_+^2+a^2\cos^2\!\theta$.
Using \eqref{temp1} we obtain
\be\label{temp44}
T=\frac{8\pi \rho_+^4 T^r{}_r|_{r_+} +r_+^2-a^2+3r_+^4/l^2+r_+^2a^2/l^2}{4\pi r_+(r_+^2+a^2)}\,,
\ee
which yields
\be\label{ho}
T\delta S=\frac{2\rho_+^4 T^r{}_r|_{r_+}}{r_+(r_+^2+a^2)}\delta S+\frac{r_+^2-a^2+\frac{3r_+^4}{l^2}+\frac{r_+^2a^2}{l^2}}{4\pi r_+(r_+^2+a^2)}\delta S
\ee
upon multiplying by $\delta S$. The final step is to realize that the latter term can be re-written as  $\delta E-\Omega\delta J$, in terms of `Kerr-AdS' quantities
\be\label{EJ2}
E=\frac{(r_+^2+a^2)(l^2+r_+^2)}{2l^2r_+\Xi^2}\,,\quad J=aE\,.
\ee
Hence we recover the HFL
\be\label{AdS2}
\delta E=T\delta S+\Omega \delta J-\sigma \delta A\,,
\ee
where, as before
\be
\sigma=\sigma(r_+,a)=\frac{\rho_+^4T^r{}_r|_{r_+}}{2r_+(r_+^2+a^2)}\,.
\ee
Eq. \eqref{temp44} gives a modified HES which now reads
\be
\sigma=\sigma(A,J,T)=
\frac{T}{4}+\frac{a^2-r_+^2-3r_+^4/l^2-r_+^2a^2/l^2}{16\pi r_+(r_+^2+a^2)}\,,
\ee
Here $r_+$ and $a$ are implicitly given in terms of $J$ and $A$ through relations \eqref{EJ2} and \eqref{S2}.

Furthermore,  in the asymptotically AdS case one can extend HFL \eqref{AdS2} to include variations of the cosmological constant, obtaining so a cohomogeneity-3 relation. Namely, upon identifying the cosmological pressure as
\be
P_\Lambda=-\frac{\Lambda}{8\pi}=\frac{3}{8\pi l^2}\,,
\ee
and varying $l$ in all the above expressions, we recover an extended HFL
\be
\delta E=T\delta S+\Omega \delta J-\sigma \delta A+{\cal V} \delta P_\Lambda\,,
\ee
where
\be
{\cal V}=\frac{r_+A}{3}\Bigl(1+\frac{1+r_+^2/l^2}{2r_+^2}\frac{a^2}{\Xi}\Bigr)
\ee
is the thermodynamic volume of Kerr-AdS black hole, e.g. \cite{Altamirano:2014tva}. An alternative derivation of this expression ala previous appendix is of course also possible.

{We finally note that the above derivation can be repeated for asymptotically de Sitter spacetimes by everywhere reversing the sign of $l^2$.  This will yield an HFL for the de Sitter black hole horizon.  An analogous HFL for the cosmological horizon remains to be understood.}

\section{Tautological argument for recovering the standard first law}

It has been claimed in  \cite{Kothawala:2007em} that for the Kerr-Newmann solution
the radial Einstein equation is equivalent to the standard
first law of black hole thermodynamics. In this appendix we briefly review this argument and comment on
its tautological character.

The argument in  \cite{Kothawala:2007em} approximately goes as follows. (In fact we describe a slightly generalized argument where variations of rotation parameter $a$ are allowed.)
One starts again with the ansatz \eqref{metric}, and identifies
 $T$, $S$, and $\Omega$ according to \eqref{temp1}, \eqref{A}, and \eqref{Omega}. The radial Einstein equation then implies
Eq.~\eqref{ho},
\be\label{app5}
T\delta S=\left(\frac{2\rho_+^4 T^r{}_r|_{r_+}}{r_+(r_+^2+a^2)}+\frac{r_+^2-a^2}{4\pi r_+(r_+^2+a^2)}\right)\delta S\,.
\ee
Next, the following specific properties of the Kerr--Newmann solution are used:
\be
 a=J/M\,,  \quad
\Delta = r^2+a^2-2mr+e^2\,,
\ee
where the latter is used to write
\be\label{rp}
r_+=M+\sqrt{M^2-(J/M)^2-Q^2}\,,
\ee
upon identifying $M=m$ and $Q=e$.
Differentiating these relations we get $\delta r_+$ and $\delta a$ in terms of variations of $\delta M, \delta J$,
and $\delta Q$.  Inserting the additional expression
\be
T^r{}_r|_{r=r_+} = -\frac{Q^2}{8\pi \rho_+^4}
\ee
(valid for the Kerr--Newman solution)  into \eqref{app5},
 one can easily verify from these relations that
\be\label{first10}
TdS=dM-\Omega dJ-\Phi dQ\,,
\ee
where $\Phi=\frac{er_+}{a^2+r_+^2}$. This concludes the proof in \cite{Kothawala:2007em} showing that the
radial Einstein equation `implies' the first law of black hole thermodynamics.

However, as obvious from this derivation, one needs to identify the correct thermodynamic charges,
\be
M=m\,,\quad J=Ma\,,\quad Q=e\,,
\ee
in order to write \eqref{rp}. Once this is known, together with identification of $T$ and $S$, one does not need to invoke the Einstein equation to write the first law \eqref{first10}. This is simply given as a `unique' linear combination of differentials of these charges.
In other words, in the above argument the radial Einstein equation is not truly needed to write the first law.

\bigskip


\begin{thebibliography}{10}

\bibitem{Padmanabhan:2002sha}
T.~Padmanabhan, {\it {Classical and quantum thermodynamics of horizons in
  spherically symmetric space-times}},  {\em Class. Quant. Grav.} {\bf 19}
  (2002) 5387--5408, [\href{http://arxiv.org/abs/gr-qc/0204019}{{\tt
  gr-qc/0204019}}].

\bibitem{Cai:2014bwa}
R.~G. Cai, {\it {Connections between gravitational dynamics and
  thermodynamics}},  {\em J. Phys. Conf. Ser.} {\bf 484} (2014) 012003.

\bibitem{Padmanabhan:2009vy}
T.~Padmanabhan, {\it {Thermodynamical Aspects of Gravity: New insights}},  {\em
  Rept. Prog. Phys.} {\bf 73} (2010) 046901,
  [\href{http://arxiv.org/abs/0911.5004}{{\tt arXiv:0911.5004}}].

\bibitem{Padmanabhan:2013xyr}
T.~Padmanabhan and D.~Kothawala, {\it {Lanczos-Lovelock models of gravity}},
  {\em Phys. Rept.} {\bf 531} (2013) 115--171,
  [\href{http://arxiv.org/abs/1302.2151}{{\tt arXiv:1302.2151}}].

\bibitem{Kothawala:2007em}
D.~Kothawala, S.~Sarkar, and T.~Padmanabhan, {\it {Einstein's equations as a
  thermodynamic identity: The Cases of stationary axisymmetric horizons and
  evolving spherically symmetric horizons}},  {\em Phys. Lett.} {\bf B652}
  (2007) 338--342, [\href{http://arxiv.org/abs/gr-qc/0701002}{{\tt
  gr-qc/0701002}}].

\bibitem{Akbar:2007zz}
M.~Akbar and A.~A. Siddiqui, {\it {Charged rotating BTZ black hole and
  thermodynamic behavior of field equations at its horizon}},  {\em Phys.
  Lett.} {\bf B656} (2007) 217--220,
  [\href{http://arxiv.org/abs/1009.3749}{{\tt arXiv:1009.3749}}].

\bibitem{Chakraborty:2015hna}
S.~Chakraborty and T.~Padmanabhan, {\it {Thermodynamical interpretation of the
  geometrical variables associated with null surfaces}},  {\em Phys. Rev.} {\bf
  D92} (2015), no.~10 104011, [\href{http://arxiv.org/abs/1508.0406}{{\tt
  arXiv:1508.0406}}].

\bibitem{KastorEtal:2009}
D.~Kastor, S.~Ray, and J.~Traschen, {\it {Enthalpy and the Mechanics of AdS
  Black Holes}},  {\em Class.Quant.Grav.} {\bf 26} (2009) 195011,
  [\href{http://arxiv.org/abs/0904.2765}{{\tt arXiv:0904.2765}}].

\bibitem{Dolan:2014jva}
B.~P. Dolan, {\it {Black holes and Boyle's law — The thermodynamics of the
  cosmological constant}},  {\em Mod. Phys. Lett.} {\bf A30} (2015), no.~03n04
  1540002, [\href{http://arxiv.org/abs/1408.4023}{{\tt arXiv:1408.4023}}].

\bibitem{Kubiznak:2014zwa}
D.~Kubiznak and R.~B. Mann, {\it {Black hole chemistry}},  {\em Can. J. Phys.}
  {\bf 93} (2015), no.~9 999--1002, [\href{http://arxiv.org/abs/1404.2126}{{\tt
  arXiv:1404.2126}}].

\bibitem{Altamirano:2014tva}
N.~Altamirano, D.~Kubiznak, R.~B. Mann, and Z.~Sherkatghanad, {\it
  {Thermodynamics of rotating black holes and black rings: phase transitions
  and thermodynamic volume}},  {\em Galaxies} {\bf 2} (2014) 89--159,
  [\href{http://arxiv.org/abs/1401.2586}{{\tt arXiv:1401.2586}}].

\bibitem{Ma:2015llh}
M.-S. Ma and R.~Zhao, {\it {Stability of black holes based on horizon
  thermodynamics}},  {\em Phys. Lett.} {\bf B751} (2015) 278--283,
  [\href{http://arxiv.org/abs/1511.03508}{{\tt arXiv:1511.03508}}].

\bibitem{Hansen:2016ayo}
D.~Hansen, D.~Kubiznak, and R.~B. Mann, {\it {Universality of P-V Criticality
  in Horizon Thermodynamics}},  \href{http://arxiv.org/abs/1603.05689}{{\tt
  arXiv:1603.05689}}.

\bibitem{Medved:2004tp}
A.~J.~M. Medved, D.~Martin, and M.~Visser, {\it {Dirty black holes: Symmetries
  at stationary nonstatic horizons}},  {\em Phys. Rev.} {\bf D70} (2004)
  024009, [\href{http://arxiv.org/abs/gr-qc/0403026}{{\tt gr-qc/0403026}}].


\bibitem{York:1986it}
  J.~W.~York, Jr.,
  {\em Black hole thermodynamics and the Euclidean Einstein action},
  {\em Phys. Rev.}  {\bf D33} (1986), 2092.





\bibitem{Herdeiro:2015waa}
C.~A.~R. Herdeiro and E.~Radu, {\it {Asymptotically flat black holes with
  scalar hair: a review}},  {\em Int. J. Mod. Phys.} {\bf D24} (2015), no.~09
  1542014, [\href{http://arxiv.org/abs/1504.0820}{{\tt arXiv:1504.0820}}].

\bibitem{Armas:2015qsv}
J.~Armas, N.~A. Obers, and M.~Sanchioni, {\it {Gravitational Tension, Spacetime
  Pressure and Black Hole Volume}},  \href{http://arxiv.org/abs/1512.09106}{{\tt
  arXiv:1512.09106}}.

\bibitem{Cvetic:2010jb}
  M.~Cvetic, G.~W.~Gibbons, D.~Kubiznak and C.~N.~Pope,
  {\em Black Hole Enthalpy and an Entropy Inequality for the Thermodynamic Volume},
  {\em Phys. Rev.} {\bf D84} (2011) 024037,
  [\href{http://arxiv.org/abs/1012.2888}{{\tt arXiv:1012.2888}}].
  %


\end{thebibliography}

\providecommand{\href}[2]{#2}\begingroup\raggedright\endgroup

\end{document}